%% file: sbaconf.tex
\documentclass[a4paper]{ifacconf}

\usepackage{graphicx,amsmath,url}      
\usepackage[round]{natbib}             

\ifx\portugues\undefined
\def\portugues{0}
\fi

\if\portugues0
   \usepackage[english]{babel}
  \else
   \usepackage[spanish,brazil,english]{babel}
\fi

\usepackage[T1]{fontenc}

\usepackage[utf8]{inputenc}

\usepackage{ae}

\usepackage{graphicx}
\usepackage{url}
\usepackage{booktabs}
\usepackage{silence}
\usepackage{caption}
\usepackage{subcaption}
\usepackage{multicol}
\usepackage{xcolor}

\usepackage{amsmath}
\usepackage{amssymb}
\usepackage{amsfonts}
\usepackage{stfloats}
\usepackage{makecell}
\usepackage{comment}
\usepackage{wrapfig}
\usepackage{xstring}  
\usepackage{forest}
\usetikzlibrary{shapes.geometric}

\newcommand{\btroot}{$\bullet$}
\newcommand{\sequence}{$\rightarrow$}
\newcommand{\fallback}{$?$}
\newcommand{\reactivesequence}{$\Rightarrow$}

\forestset{
  controlflow/.style={
    draw,
    rectangle,
    rounded corners=2pt,
    minimum width=1.4cm,
    minimum height=0.65cm,
    align=center,
    font=\small\bfseries,
    fill=white,
    edge={-latex},
  },
  subtree/.style={
    draw,
    dashed,
    rectangle,
    minimum width=2.2cm,
    minimum height=0.65cm,
    align=center,
    font=\small,
    fill=white!90!gray,
    edge={-latex},
  },
  condition/.style={
    draw,
    ellipse,
    minimum width=2.2cm,
    minimum height=0.65cm,
    align=center,
    font=\small\itshape,
    fill=white,
    edge={-latex},
  },
  action/.style={
    draw,
    rectangle,
    minimum width=2.2cm,
    minimum height=0.65cm,
    align=center,
    font=\small,
    fill=white,
    edge={-latex},
  },
}

\newif\ifanonymous
\anonymousfalse   

\newcommand{\authoranon}[1]{%
  \ifanonymous
    \author{Anonymous Author(s)}%
  \else
    #1%
  \fi
}

\newcommand{\textanon}[2]{%
  \ifanonymous
    #2%
  \else
    #1%
  \fi
}

\newcommand{\anontext}{<text removed for peer review>}

\newcommand{\addressanon}[2]{%
  \ifanonymous
    \address[#1]{\anontext}%
  \else
    \address[#1]{#2}%
  \fi
}

\newcommand{\citeanon}[2]{%
  \ifanonymous
    #2%
  \else
    #1%
  \fi
}

\newcommand{\linkanon}[2]{%
  \ifanonymous
    #2%
  \else
    #1%
  \fi
}

\if\portugues1
 { }
 { }
 { }
 
\fi

\begin{document}

\if\portugues1

%
\selectlanguage{brazil}
	
\begin{frontmatter}

\title{Estilo para Artigos das Conferências da SBA --- Adaptado do  IFAC\thanksref{footnoteinfo}} 

\thanks[footnoteinfo]{Reconhecimento do suporte financeiro deve vir nesta nota de rodapé.}

\author[First]{Primeiro A. Autor} 
\author[Second]{Segundo B. Autor} 
\author[Third]{Terceiro C. Autor}

\address[First]{Faculdade de Engenharia Elétrica, Universidade do Triângulo, MG, (e-mail: autor1@faceg@univt.br).}
\address[Second]{Faculdade de Engenharia de Controle \& Automação, Universidade do Futuro, RJ (e-mail: autor2@feca.unifutu.rj)}
\address[Third]{Electrical Engineering Department, 
   Seoul National University, Seoul, Korea, (e-mail: author3@snu.ac.kr)}

\selectlanguage{english}
\renewcommand{\abstractname}{{\bf Abstract:~}}
\begin{abstract}                
These instructions give you guidelines for preparing papers for the Sociedade Brasileira de Automática (SBA) technical meetings using the IFAC style. Please use this document as a template to prepare your manuscript in portuguese. For submission guidelines, follow instructions on paper submission system as well as the event website.

\vskip 1mm
\selectlanguage{brazil}
{\noindent \bf Resumo}:  As instruções abaixo são linhas gerais para a preparação de artigos para conferências e simpósios da Sociedade Brasileira de Automática (SBA) usando como base o estilo IFAC. Instruções de submissão podem ser encontradas no sistema de submissão de artigos ou no {\em website} do congresso.
\end{abstract}

\selectlanguage{english}

\begin{keyword}
Five to ten keywords separatety by semicolon. 

\vskip 1mm
\selectlanguage{brazil}
{\noindent\it Palavras-chaves:} Utilize de cinco a dez palavras-chaves separadas por ponto e vírgula.
\end{keyword}

\selectlanguage{brazil}

\end{frontmatter}
\else
%

\begin{frontmatter}

\title{Coordinating Task Switching in a Robotics Multi-Agent System Using Behavior Trees\thanksref{footnoteinfo}} 

\thanks[footnoteinfo]{\textanon{This study was financed in part by the Coordenação de Aperfeiçoamento de Pessoal de Nível Superior - Brazil (CAPES) - Finance Code 001}{\anontext}}

\authoranon{%
  \author[First]{Lucas Haug} 
  \author[First]{Anarosa Alves Franco Brandão} 
  \author[First]{Arthur Casals}

  \addressanon{First}{LTI - Laboratório de Técnicas Inteligentes, Universidade de S$\tilde{a}$o Paulo, SP}

}

\textit{Preprint of a manuscript submitted to the XXVI Congresso Brasileiro de Automática (CBA 2026).}

\renewcommand{\abstractname}{{\bf Abstract:~}}   
   
\input{contents/abstract}

\begin{keyword}
Behavior trees; Multi-agent systems; Multi-vehicle and multi-robot systems; Coordination and control methods; Discrete event and hybrid systems; Robotics and autonomous systems; Control architectures
\end{keyword}

\end{frontmatter}
\fi


\input{contents/introduction}

\input{contents/related_work}

\input{contents/team_organization}

\input{contents/implementation}

\input{contents/evaluation}

\input{contents/conclusion}

\input{contents/acknowledgments}

\bibliography{ifacconf}                                                                 
                                                   
\end{document}

%% file: contents/abstract.tex
\begin{abstract}

The application of multi-agent systems in robotics is a very challenging field. Several competitions involving such systems are proposed to foster research and development of strategies and mechanisms using games as the underlying domain. Among them are the ones from the \textit{IEEE Very Small Soccer (VSSS)} category, which is the case study described in this paper. In VSSS, two teams of three robots each compete in a very dynamic environment of a soccer game. Thus, coordination of robots' behavior during the game is crucial to win it. 
In this paper, we present a Behavior-Tree-based approach to support multi-robot coordination within the VSSS team of the \textanon{ThundeRatz}{\anontext} robotics team from the \textanon{Universidade de S$\tilde{a}$o Paulo}{\anontext}. Moreover, a comparison between the proposed approach and the previous one, which was based on a Finite State Machine (FSM), was conducted using the FIRASim simulator. Besides that, the performance of this new strategy was further evaluated in an academic robotics competition.

\end{abstract}

%% file: contents/introduction.tex
\section{Introduction}

\def \MOISEp {$\mathcal{M}OISE^+$} 

One of the biggest challenges in multi-agent systems is providing a way to coordinate multiple agents to achieve a common goal. These challenges encompass both the coordination logic, as the ones described in \cite{DBLP:journals/dagstuhl-reports/AgotnesB14}, and the control architecture (e.g., Finite State Machines, Behavior Trees, Decision Trees) used to implement the defined logic, which depend on the architecture of the system being coordinated. To successfully coordinate agents, it is crucial to have a well-defined coordination logic and an appropriate control architecture that is compatible with the system architecture.

In this regard, the \citeanon{ThundeRatz~\cite{ThundeRatz}}{[self-citation removed for review]} robotics team at the \textanon{Universidade de S$\tilde{a}$o Paulo}{\anontext} developed a project, called \citeanon{ThunderVolt~\cite{ThunderVolt}}{[self-citation removed for review]}, to participate in academic competitions of the \textit{IEEE Very Small Size Soccer (VSSS)} \cite{VSSS} category, which has a great challenge in terms of coordinating multiple robots. In this category, three robots, the size of a cube of 75 mm on each side, play soccer against three other robots, each game lasting ten minutes. Then, to control their robots, each team uses a camera located at the top of the field and connected to a central team computer, which performs image processing to determine the states of the robots, and uses these states to define what each robot should do based on a particular coordination strategy, sending commands to the robots accordingly.

In the \textanon{ThunderVolt}{\anontext} project, to coordinate which robot should play which role in the game, the control logic was first defined based on the system architecture specified by the category. In this way, the coordination of robots is entirely based on a central computer that determines the roles of each of the agents in the organization, taking away the possibility of the robots to participate in the coordination effort, since they do not have an overview of the state of the game.

As for the implementation, the project previously used Finite State Machines (FSM) as the control architecture in the robot coordination system. However, with several developments in the project, it was noticed that the use of FSM did not scale well. In fact, such project decisions made the system much more complex to understand, difficult to maintain, and to make improvements.

For these reasons, it was necessary to update the control architecture used in the coordination system, so that an improvement in the project could be possible. After some analyses, Behavior Trees~\cite{BTsMultRobot} were chosen, as they are much more modular, flexible, and easy to understand.

This paper is structured into six sections. The second section of this paper discusses the related work, describing the relevance of this work in relation to other research. The third section introduces the \textanon{ThunderVolt}{\anontext} team organization, presenting the organization modeling and the previous coordination strategy. The fourth section details the proposed solution to enhance the team coordination strategy using BTs. The fifth section outlines the evaluation of the proposed solution. Lastly, the sixth section presents the conclusions of this work.

%% file: contents/related_work.tex
\section{Related Work}

The focus of this work is the development of a new control strategy for a multi-agent organization using classical artificial intelligence, comparing different control architectures. Given these restrictions, machine learning methods are not used, even though they can be integrated with the control architectures described below, for example, in \cite{LearningBTs,LearningFSMs}.

There is extensive research on the field of control architectures for robotics systems \cite{BTsInRobotics,SurveyBTs,Expressiveness,iovino2022programming,RobotArchitectureInDynamicEnvironment,PetriNetsRobotics,billington2010plausible,BTsAndFSMApplications}, demonstrating the use of different control architectures and comparing them. Most of the systems in robotics tend to use Finite State Machines (FSMs) and Hierarchical Finite State Machines (HFSMs) due to their simplicity, comprehensibility, maturity, and widespread use \cite{BTsInRobotics,iovino2022programming,BTsAndFSMApplications}. However, Behavior Trees (BTs) have gained more attention in recent years and are now almost as widely used as FSMs and HFSMs \cite{BTsAndFSMApplications}.

Even though FSMs and HFSMs are very popular, when implementing more complex applications, it is possible to observe how systems using these architectures scale poorly, complicating their maintenance, as shown in \cite{BTsInRobotics,SurveyBTs,iovino2022programming}. In contrast, BTs offer expressiveness, modularity, and flexibility, generalizing many other control architectures, such as the Subsumption Architecture, the Teleo-Reactive Paradigm, and Decision Trees \cite{BTsInRobotics}. In terms of expressiveness, HFSMs are the most similar control architecture to BTs, offering a powerful and robust architecture. However, as previously mentioned, HFSMs also have their disadvantages. For this reason, BTs were chosen as the control architecture used to implement the new control strategy of the team, enabling greater scalability, modularity, and ease of maintenance.

Nevertheless, in most of these researches, the focus is more on modeling the individual behaviors of isolated intelligent agents, while comparing control architectures. Concerning the use of BTs in multi-agent systems, there are other papers that are more focused on this area, such as \cite{Event-DrivenBTs}, from the game development field, and \cite{Self-ReactivePlanningOfMulti-Robots,BTsMultRobot}, from the robotics fields. Agis, Gottifredi, and García~\cite{Event-DrivenBTs} propose an extension of BTs for distributed multi-agent coordination in games, introducing a framework that enables agents to react to events and communicate with each other. Meanwhile, Yang and colleagues~\cite{Self-ReactivePlanningOfMulti-Robots}, propose the coordination of swarms of robots by using a BT that formalizes a coordination behavior and applies it to all robots, enabling information exchange among them to perform tasks. Nonetheless, the article does not detail how to handle the task assignment part using a BT. Lastly, Colledanchise and colleagues~\cite{BTsMultRobot} highlight the benefits of employing BTs in multi-robot systems, demonstrating a system that is composed of two types of trees, one responsible for performing tasks and the other responsible for managing the task assignment. However, it does not address how to handle task switches between the robots.

Therefore, as the \textanon{ThunderVolt}{\anontext} project necessitates continuous and real-time responsiveness, dynamic task assignment, and task switching between robots, it serves as a valuable case study for exploring the use of BTs in multi-agent organizations, extending the research conducted in existing works.

%% file: contents/team_organization.tex
\def \MOISEp {$\mathcal{M}OISE^+$ } 
\def \MOISEpBf {$\mathbf{\mathcal{M}OISE^+}$ } 

\section{Team Organization}

As specified before, the \textanon{ThunderVolt}{\anontext} project is a project implemented by the \textanon{ThundeRatz}{\anontext} robotics team, which competes in the VSSS category. The team's primary objective is to score as many goals as possible against their opponents. To achieve this objective, the members of the \textanon{ThunderVolt}{\anontext} project have defined seven roles that the three robots of the team can play. These roles are designed to improve the team's performance in the dynamic environment of a soccer game. The seven roles are described in more detail below.

\begin{itemize}
    \item \textit{Goalkeeper}: Defends the goal.
    \item \textit{Striker}: Carries out attacks against the opposing team.
    \item \textit{Assistant}: Strategically positions itself dynamically to take advantage of rebounds in the attack state.
    \item \textit{Fullback}: Helps the goalkeeper to defend the goal
    \item \textit{Wingback}: Helps the defense, while trying to make counterattacks.
    \item \textit{Penalty Kicker}: Kicks penalties in the opposing team.
    \item \textit{Penalty Defender}: Defends the opposing team's penalties.
\end{itemize}

As there are more roles than robots it is extremely necessary to have a form of coordinating which robot should play each role. For that, we adopted the \MOISEp to model how the coordination of the team occurs. This modeling is presented in the next subsection. Regarding the coordination strategy of the team, it is carried out by the central computer used in the category and is the one that was performed by an FSM but became too hard to maintain and improve.

\subsection{Modeling}

\begin{figure*}[!ht]
    \centering
    \includegraphics[width=1\linewidth]{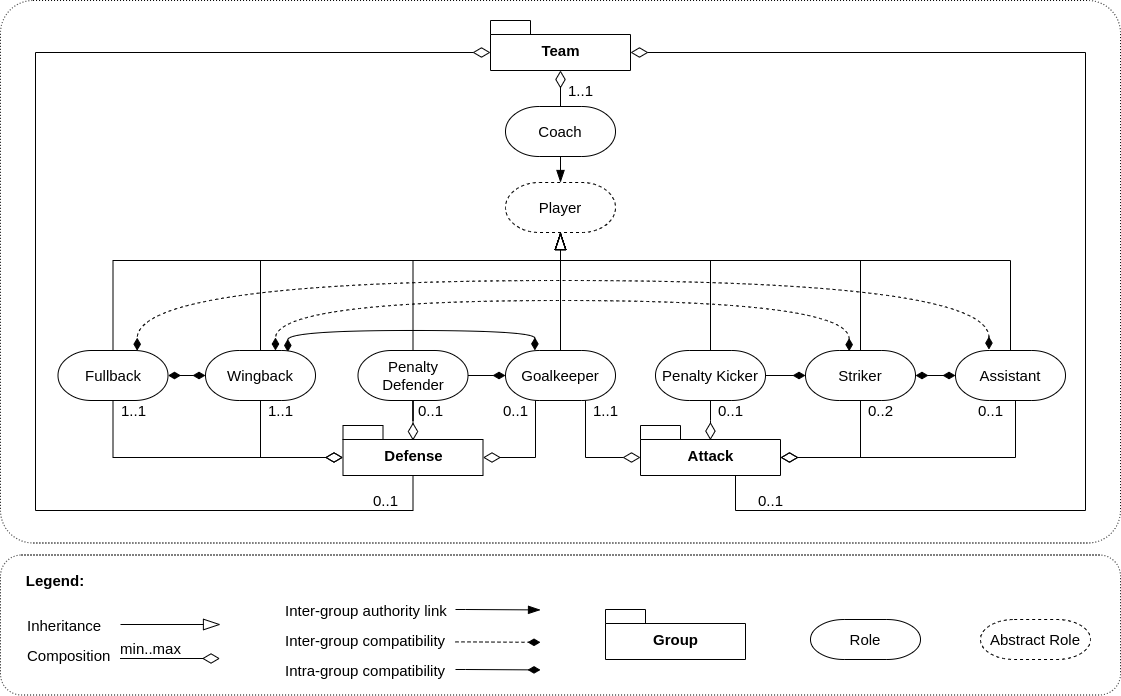}
    \caption{Mapping of the system structural specification using \MOISEpBf.}
    \label{fig:specification}
\end{figure*}

For the modeling, the \MOISEp framework was used, which was chosen due to its suitability to the application and the authors' familiarity with the model. The structural specification of the team can be seen in Figure \ref{fig:specification}. 

In the structural specification, it is possible to observe the relationship between the different roles that each robot can play, this is represented by the inter- and intra-group compatibilities. The inter-group compatibilities show the relationships between roles from the same group, which means the relationships between the roles in the defense and the relationships between the roles in the attack. For example, a robot playing the Fullback role can also play the Wingback role. The intra-group compatibilities, on the other hand, show the relationships between the roles from different groups, for example, a robot playing the Wingback role, when changing from the defense state to the attack state will play the Striker role due to their compatibility. 

It is also possible to observe how a leader role called \textbf{Coach} is defined, having authority over all the other roles. In addition, it is through this role that the coordination strategy was modeled using an FSM. The artifice of authority over other roles was used to have a global way of defining which robot should play which role, delegating that authority to an agent with a more general view of the game and therefore simplifying the system. 

In the specification, it is also defined how many robots should play a role in a defined group and how many of each group of roles should be used, which is an important information for developing the coordination strategy of the team. These definitions are specified as the compositions. For example, in the Attack group, there will always be a robot playing the Goalkeeper role, however, it is possible to have different combinations of the other roles, e.g. two Strikers or one Striker and one Assistant.

\subsection{Previous coordination solution}

\begin{figure*}[!ht]
    \centering
    \includegraphics[width=1.05\linewidth]{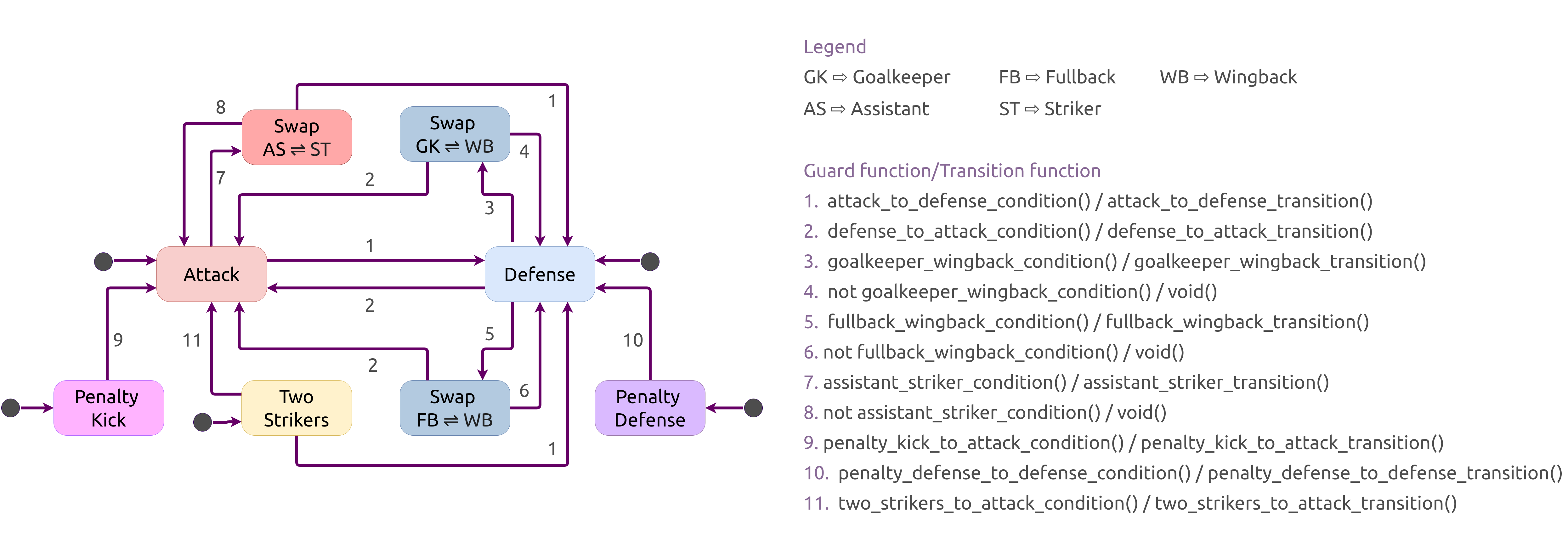}
    \caption{Previous FSM of the coordination strategy.}
    \label{fig:behaviors_controller_fsm}
\end{figure*}

The previous software solution for the robotics team used an FSM, which can be seen in Figure \ref{fig:behaviors_controller_fsm}. The FSM comprises eight possible states, consisting of three states involving role swaps and five input states. The input state defines the initial configuration of the roles that are being used and the swap states perform the swaps between the roles. 

The state in which the state machine initiates is determined beforehand upon receiving an event from an automated referee developed for the category \cite{VSSReferee}, so all the initialization of the strategy is done before starting the FSM. An example would be receiving a penalty event for the opposing team, whereby the entry state would be the penalty defense state and, for this case, the roles that would be used are the Penalty Defender, the Fullback, and the Wingback. Each state transition is governed by a guard function, and when the function's condition is fulfilled, a transition function is invoked. These functions are listed for each transition in Figure \ref{fig:behaviors_controller_fsm}. 

It is worth noting that, in all cases, to enter a swap state, a swap condition is assessed. Meanwhile, to exit these states and transit back to the attack or defense states, the negation of the swap condition is analyzed. This feature is crucial to ensure that the swap state is exited only when the swap is fully completed and cannot take place immediately afterward, guaranteeing a hysteresis effect on the system and avoiding multiple successive swaps that would destabilize the system.

%% file: contents/implementation.tex
\def \MOISEp {$\mathcal{M}OISE^+$ } 
\def \MOISEpBf {$\mathbf{\mathcal{M}OISE^+}$ } 

\section{Proposed coordination solution using Behavior Trees}

In the proposed solution, the same \MOISEp model, as previously presented, is used and its implementation is made using Behavior Trees to improve the system coordination method. As the system already had a leader agent responsible for defining the robots' roles, it was possible to just refactor the agent that used the FSM to use a BT instead. 

In this way, modeling the team's coordination behavior is reduced to modeling the Coach's behavior. However, it is important to note that, for this improvement, the goal of the changes was not necessarily to impact at first the system's performance, but just to improve its structure and maintainability.


\begin{figure}[!t]
    \centering
    \scalebox{.8} {
        \begin{forest}
            [\btroot, controlflow
                [\sequence, controlflow  
                    [{Roles Swapper \\Initializer Subtree}, subtree]
                    [{Roles Swapper \\Subtree}, subtree]
                ]
            ]
        \end{forest}
    }
    \caption{Base structure of the Coach’s Behavior Tree.}
    \label{fig:coach_bt}
\end{figure}
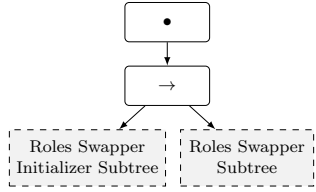


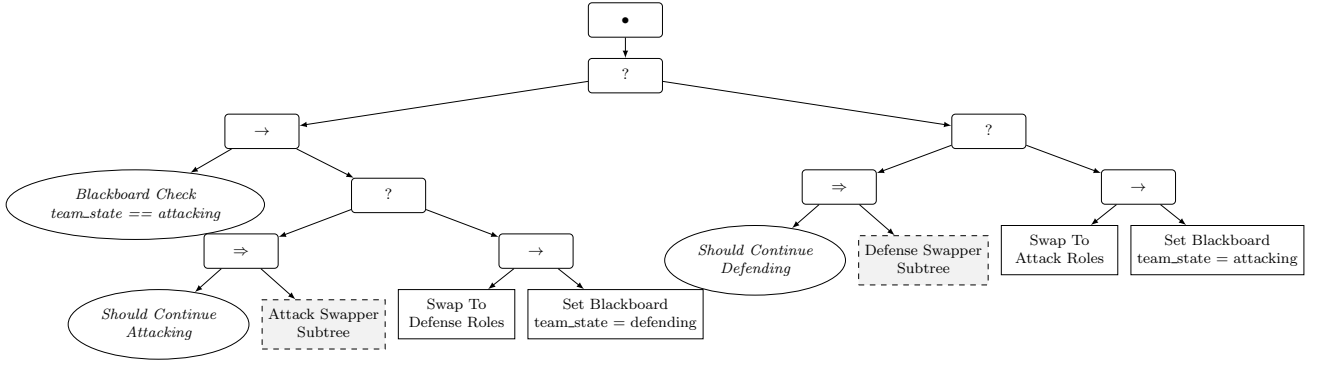
\begin{figure*}[!t]
    \centering
    \scalebox{.7} {
        \begin{forest}
            [\btroot, controlflow
                [\fallback, controlflow    
                    [\sequence, controlflow      
                        [{Blackboard Check \\ team\_state == attacking}, condition]
                        [\fallback, controlflow        
                            [\reactivesequence, controlflow          
                                [{Should Continue \\Attacking}, condition]
                                [{Attack Swapper \\Subtree}, subtree]
                            ]
                            [\sequence, controlflow 
                                [{Swap To \\Defense Roles}, action]
                                [{Set Blackboard \\ team\_state = defending}, action]
                            ]
                        ]
                    ]
                    [\fallback, controlflow        
                        [\reactivesequence, controlflow          
                            [{Should Continue \\Defending}, condition]
                            [{Defense Swapper \\Subtree}, subtree]
                        ]
                        [\sequence, controlflow 
                            [{Swap To \\Attack Roles}, action]
                            [{Set Blackboard \\ team\_state = attacking}, action]
                        ]
                    ]
                ]
            ]
        \end{forest}
    }
    \caption{Roles Swapper Subtree.}
    \label{fig:roles_swapper}
\end{figure*}


\begin{figure*}[!t]
    \centering
    \scalebox{.75} {
        \begin{forest}
            [\btroot, controlflow
                [\fallback, controlflow    
                    [\sequence, controlflow      
                        [{Blackboard Check \\use\_two\_strikers\_mode == True}, condition]
                        [\sequence, controlflow        
                            [{Change from Two Strikers \\to One Striker mode}, action]
                            [{Set Blackboard \\use\_two\_strikers\_mode = False}, action]
                        ]
                    ]
                    [\sequence, controlflow      
                        [{Blackboard Check \\use\_penalty\_mode == True}, condition]
                        [\sequence, controlflow        
                            [{Change Penalty \\Kicker to Striker}, action]
                            [{Set Blackboard \\use\_penalty\_mode = False}, action]
                        ]
                    ]
                    [{Swap Striker \\and Assistant}, action]
                    [{Always Success}, action]
                ]
            ]
        \end{forest}
    }
    \caption{Attack state internal swap subtree.}
    \label{fig:attack_swapper}
\end{figure*}
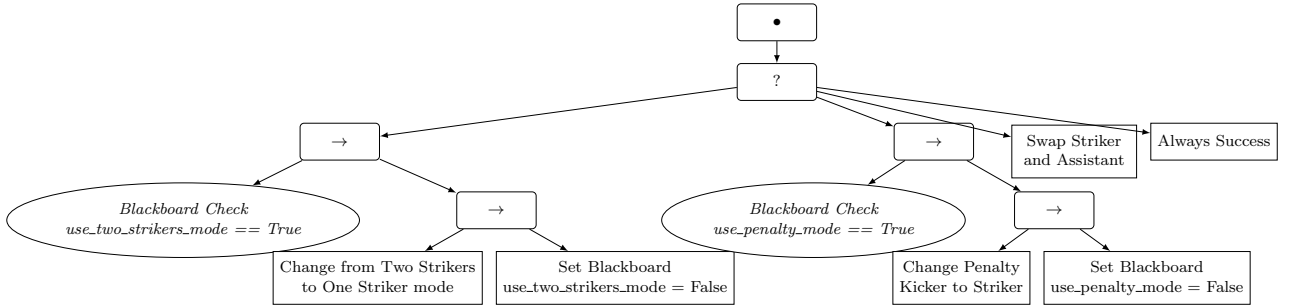


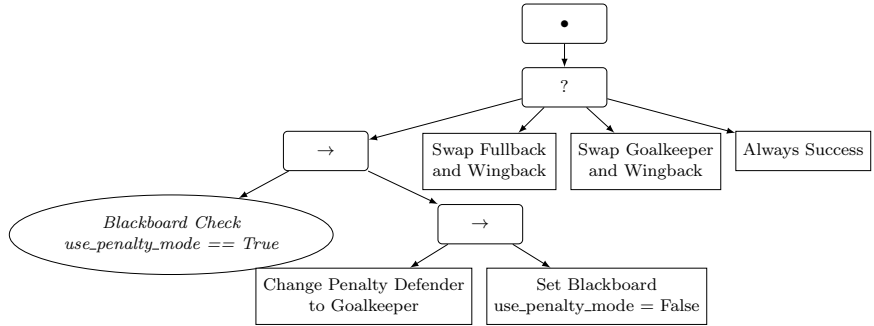
\begin{figure*}[!t]
    \centering
    \scalebox{.8} {
        \begin{forest}
            [\btroot, controlflow
                [\fallback, controlflow    
                    [\sequence, controlflow      
                        [{Blackboard Check \\use\_penalty\_mode == True}, condition]
                        [\sequence, controlflow        
                            [{Change Penalty Defender \\to Goalkeeper}, action]
                            [{Set Blackboard \\use\_penalty\_mode = False}, action]
                        ]
                    ]
                    [{Swap Fullback \\and Wingback}, action]
                    [{Swap Goalkeeper \\and Wingback}, action]
                    [{Always Success}, action]
                ]
            ]
        \end{forest}
    }
    \caption{Defense state internal swap subtree.}
    \label{fig:defense_swapper}
\end{figure*}

For the case of the application, a tree with two main branches was developed, as can be seen in Figure \ref{fig:coach_bt}. The left branch takes care of initializing the roles of each robot after each pause in the game, this branch is responsible for adapting the coordination system to the rules of the category \cite{RulesVSSS}, which the system has to obey. Also, it handles messages received from an automatic referee system \cite{VSSReferee} reporting what happened in the game and depending on what happened, sets different starting roles for each robot. This branch is omitted here for simplification, as it just handles the message events. This initialization subtree replaces the functionality of several FSM initial states, making the strategy more integrated and simpler to understand.

The right branch is responsible for performing the dynamic analysis of role changes and its structure is depicted in Figure \ref{fig:roles_swapper}. It has two sub-branches, one for the moment when the team is attacking (left branch of the fallback in the root) and the other for when it is defending (right branch of the fallback in the root). This subtree is the one responsible for performing the role changes between the roles from the attack group and the ones from the defense group. The idea of the two sub-branches follows a principle very similar to the two states of attack and defense present in the previous FSM-based strategy.

The tracking of the state of the team, whether it is attacking or defending, is done using a blackboard \cite{BlackboardDesignPattern}, a structure used to share environment variables between nodes. Then, in the \textit{Roles Swapper Initializer} subtree, it is possible to easily set in the blackboard if the team should attack or defend, and in the \textit{Roles Swapper} subtree to swap the state of the team by changing the same variable.

The two sub-branches of the \textit{Roles Swapper} subtree have very similar structures, with the main difference being the subtrees that each branch includes. The attack branch includes the \textit{Attack Swapper} subtree (Figure \ref{fig:attack_swapper}), which handles switching between roles that are part of the attack role group (see Figure \ref{fig:specification}), and the \textit{Defense Swapper} subtree (Figure \ref{fig:defense_swapper}) handles switching between roles in the defense group. The \textit{Attack Swapper} and \textit{Defense Swapper} subtrees encompass the functionalities of the FSM's role swaps states.

The \textit{AttackSwapper} subtree first handles when the team is using a mode where two Strikers roles are played by two robots at the same time, then it handles the case where the team is kicking a penalty. Finally, it checks if the robot playing the Striker role needs to swap roles with the robot playing the Assistant role, and, if nothing needs to be done, it returns success. The \textit{DefenseSwapper} subtree has similar behavior, it first handles the case when the team is defending a penalty kick, then it handles the case of swapping the roles of the Fullback and the Wingback, then the case of swapping the Goalkeeper and the Wingback, and, if nothing needs to be done, it also returns success.

To validate the proposed model for the coordinating agent, the BT was implemented using the \textit{BehaviorTree.CPP} \cite{BehaviorTree.CPP} library, which was chosen for its open-source nature, large community, and wide use in different robotics applications.

%% file: contents/evaluation.tex
\section{Evaluation}

\subsection{Functional and non-functional comparison}

To make a comparison between the FSM-based team and the BT-based team, it is necessary to take into account the objectives of the refactoring. The change was mainly aimed at modifying the system's control architecture to improve its maintenance, its understanding, and its flexibility to change, but without necessarily impacting the team's performance at first.

It is possible to state that the system's maintainability and flexibility improved since the system became more modular and scalable. The system increased its modularity thanks to the independent nodes that were developed in the tree and thanks to the division of the system into subtrees. Besides that, it became more scalable since to add new functionalities, such as, for example, new roles in the organization, it is not necessary to worry about numerous transitions between states. In fact, in the case of adding new roles, it is only necessary to add an extra node that takes care of the swap between the new role and another compatible one.

Regarding the ease of understanding, it is possible to state that the graphical representation of the BT may seem more complex than that of the FSM, but this is because the structure developed for the BT is much more transparent of its functionalities than that of the FSM. A clear example of this fact is the priority between role changes, a characteristic that is not possible to see clearly only with the graphic representation of the FSM, but is quite explicit in the BT.

\subsection{Tests between the FSM-based and BT-based teams}

\begin{figure}[!t]
    \centering
    \resizebox{0.46\textwidth}{!}{
        \input{images/goals_diff_histogram.pgf}
    }
    \caption{Goals difference histogram: $\mu = 0.2120, \sigma = 1.1794$.}
    \label{fig:goals_diff_hist}
\end{figure}
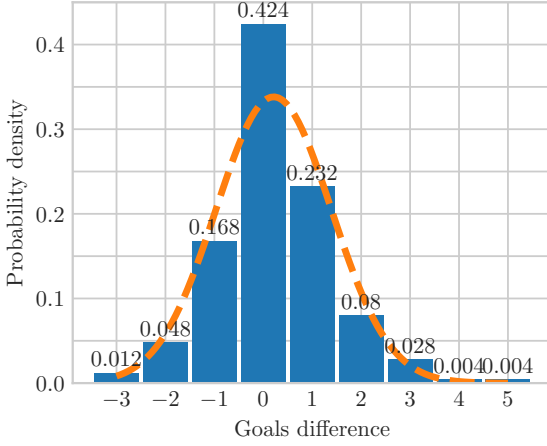

The tests were carried out using the FIRASim simulator \cite{FIRASim} and the automatic referee of the category. They consisted of 250 games of ten minutes each, totaling 2500 minutes of game. The overall results of the tests are presented in Tables \ref{tab:wins} and \ref{tab:goals_number_metrics} and in Figure \ref{fig:goals_diff_hist}.

\begin{table*}[!t]
    \caption{Evaluation metrics}
    \begin{subtable}[ht]{\columnwidth}
        \centering
        \caption{Wins, losses, and \\ties ratio.}
        \label{tab:wins}
        \begin{tabular}{c c}
            \toprule
            Wins and ties       & Ratio \\
            \midrule
            \makecell{BT-based \\Team Wins}  & 34.80\% \\
             Ties                & 42.40\% \\
            \makecell{FSM-based \\Team Wins} & 22.80\% \\
            \bottomrule
        \end{tabular}
    \end{subtable}
    \hfill
    \begin{subtable}[ht]{\columnwidth}
        \centering
        \caption{Goals related metrics.}
        \label{tab:goals_number_metrics} 
        \begin{minipage}{\columnwidth}
            \begin{center}
                \begin{tabular}{l c c}
                    \toprule
                    Teams                     & BT-based & FSM-based \\
                    \midrule
                    Total goal                & 193      & 140       \\
                    Maximum goals in a game   & 5        & 3         \\
                    GPG - average             & 0.7720   & 0.5600    \\
                    GPG - standard deviation  & 0.9550   & 0.7526    \\
                    GDPG - average            & 0.2120   & -0.2120   \\
                    GDPG - standard deviation & 1.1794   & 1.1794    \\
                    \bottomrule
                \end{tabular}
            \end{center}    
        \end{minipage}
    \end{subtable}
    \vspace{4pt}
    \begin{center}
        \footnotesize 
        \emph{Legend:} GPG: Goals per game; GDPG: Goals difference per game.\\   
    \end{center}
\end{table*}

From Table \ref{tab:wins}, it is possible to note that, although most of the games between the two teams were ties, the BT-based team had a higher percentage of wins than the FSM-based team. Furthermore, Table \ref{tab:goals_number_metrics} indicates that the BT-based team had a superior performance regarding the number of goals (193 to 140 - or 35\% more efficient considering absolute numbers). In addition, the average number of goals per game was also higher for the BT-based team (0.7720 to 0.5600 - almost 23\% higher). 

Considering the goals difference per game metric, it is possible to perform a one-sample t-test, taking into account the null hypothesis that the two teams have similar performance. As for the null hypothesis, the average goals difference per game should be zero, the p-value calculated for this case is 0.00243, which makes it possible to reject the null hypothesis with a significance of 5\%, showing that indeed the BT-based team performed better than the FSM-based team.

This performance improvement is probably due to better control over the strategy reactivity and better control of certain characteristics of the strategy, such as the priority between roles changes.

\subsection{Robotics Academic Competition}

\def \thundervolt{\textanon{ThunderVolt}{Target Team}}
\def \robocin{\textanon{RobôCIn}{Team A}}
\def \robotbulls{\textanon{Robotbulls}{Team B}}
\def \itandroids{\textanon{ITAndroids}{Team C}}
\def \reddragons{\textanon{Red Dragons}{Team D}}
\def \rinobot{\textanon{Rinobot}{Team E}}
\def \neon{\textanon{Neon}{Team F}}

The implementation was also tested in a Brazilian robotics academic competition, the \textanon{IRONCup 2023}{\anontext} \citeanon{IRONCup2023}. The competition consisted of a round-robin tournament and was played in the virtual environment of the FIRASim simulator between the following teams: \textanon{ThunderVolt}{target team of this paper}, \robocin\footnote{\linkanon{https://robocin.com.br/}{https://robocin.com.br/}}, \robotbulls\footnote{\linkanon{https://inatel.br/robotica/}{https://inatel.br/robotica/}}, \itandroids\footnote{\linkanon{https://www.itandroids.com.br/en/}{https://www.itandroids.com.br/en/}}, \reddragons\footnote{\linkanon{https://www.linkedin.com/company/reddragons/}{https://www.linkedin.com/company/reddragons/}}, \rinobot\footnote{\linkanon{https://www.linkedin.com/company/rinobot-team/}{https://www.linkedin.com/company/rinobot-team/}} and \neon\footnote{\linkanon{https://projectneon.dev/}{https://projectneon.dev/}}.

The team won four in five games (see Table \ref{tab:ironcup_games}. The team won second place in the competition, showing the effectiveness of the coordination structure (see Table \ref{tab:ironcup_results}).

\begin{table}[!t]
    \caption{Results of the competition \cite{ResultsIRONCup2023}.}
    \begin{minipage}{\columnwidth}
    \begin{subtable}[t]{\textwidth}
        \centering
        \caption{Games played by the \thundervolt team.}
        \label{tab:ironcup_games}
        \begin{tabular}{c c c c c}
            \toprule
            Blue team &&&& Yellow team \\
            \midrule
            \thundervolt & 7 & x &  0 & \reddragons  \\
            \neon        & 0 & x & 14 & \thundervolt \\
            \thundervolt & 4 & x &  0 & \itandroids  \\
            \rinobot     & 0 & x &  5 & \thundervolt \\
            \robocin	 & 3 & x &  0 & \thundervolt \\
            \robotbulls  & 0 & x &  2 & \thundervolt \\
            \bottomrule
        \end{tabular}
    \end{subtable}
    \hfill
    \begin{subtable}[t]{\textwidth}
        \centering
        \caption{Overall results considering all games.}
        \label{tab:ironcup_results}
        \begin{tabular}{c c c c c c c c}
            \toprule
            \textbf{Team} & \textbf{Pts} & \textbf{GP} & \textbf{Vic} & \textbf{Def} & \textbf{GS} & \textbf{GC} & \textbf{GD} \\
            \midrule
            \robocin     & 18 & 6 & 6 & 0 & 68 & 17 &  51 \\
            \thundervolt & 15 & 6 & 5 & 1 & 32 &  3 &  29 \\
            \robotbulls  & 12 & 6 & 4 & 2 & 34 & 23 &  11 \\
            \itandroids  &  9 & 6 & 3 & 3 & 36 & 17 &  19 \\
            \reddragons  &  6 & 6 & 2 & 4 & 36 & 36 &   0 \\
            \rinobot     &  3 & 6 & 1 & 5 & 14 & 54 & -40 \\
            \neon        &  0 & 6 & 0 & 6 &  7 & 46 & -39 \\
            \bottomrule
        \end{tabular}
    \end{subtable}
        \bigskip
        \begin{center}
            \footnotesize 
            \emph{Legend:} Pts: Points; GP: Games Played; Vic: Victories; Def: Defeats;\\ GS: Goals Scored; GC: Goals Conceded; GD: Goal Difference.\\   
            \emph{Points Count}: Each victory counts as three points and each tie counts as one point.
        \end{center}
    \end{minipage}
\end{table}

%% file: images/goals_diff_histogram.pgf
\begingroup%
\makeatletter%
\begin{pgfpicture}%
\pgfpathrectangle{\pgfpointorigin}{\pgfqpoint{3.793206in}{3.035313in}}%
\pgfusepath{use as bounding box, clip}%
\begin{pgfscope}%
\pgfsetbuttcap%
\pgfsetmiterjoin%
\definecolor{currentfill}{rgb}{1.000000,1.000000,1.000000}%
\pgfsetfillcolor{currentfill}%
\pgfsetlinewidth{0.000000pt}%
\definecolor{currentstroke}{rgb}{1.000000,1.000000,1.000000}%
\pgfsetstrokecolor{currentstroke}%
\pgfsetdash{}{0pt}%
\pgfpathmoveto{\pgfqpoint{0.000000in}{0.000000in}}%
\pgfpathlineto{\pgfqpoint{3.793206in}{0.000000in}}%
\pgfpathlineto{\pgfqpoint{3.793206in}{3.035313in}}%
\pgfpathlineto{\pgfqpoint{0.000000in}{3.035313in}}%
\pgfpathlineto{\pgfqpoint{0.000000in}{0.000000in}}%
\pgfpathclose%
\pgfusepath{fill}%
\end{pgfscope}%
\begin{pgfscope}%
\pgfsetbuttcap%
\pgfsetmiterjoin%
\definecolor{currentfill}{rgb}{1.000000,1.000000,1.000000}%
\pgfsetfillcolor{currentfill}%
\pgfsetlinewidth{0.000000pt}%
\definecolor{currentstroke}{rgb}{0.000000,0.000000,0.000000}%
\pgfsetstrokecolor{currentstroke}%
\pgfsetstrokeopacity{0.000000}%
\pgfsetdash{}{0pt}%
\pgfpathmoveto{\pgfqpoint{0.474151in}{0.333884in}}%
\pgfpathlineto{\pgfqpoint{3.413885in}{0.333884in}}%
\pgfpathlineto{\pgfqpoint{3.413885in}{2.671076in}}%
\pgfpathlineto{\pgfqpoint{0.474151in}{2.671076in}}%
\pgfpathlineto{\pgfqpoint{0.474151in}{0.333884in}}%
\pgfpathclose%
\pgfusepath{fill}%
\end{pgfscope}%
\begin{pgfscope}%
\pgfpathrectangle{\pgfqpoint{0.474151in}{0.333884in}}{\pgfqpoint{2.939735in}{2.337191in}}%
\pgfusepath{clip}%
\pgfsetroundcap%
\pgfsetroundjoin%
\pgfsetlinewidth{0.803000pt}%
\definecolor{currentstroke}{rgb}{0.800000,0.800000,0.800000}%
\pgfsetstrokecolor{currentstroke}%
\pgfsetdash{}{0pt}%
\pgfpathmoveto{\pgfqpoint{0.742901in}{0.333884in}}%
\pgfpathlineto{\pgfqpoint{0.742901in}{2.671076in}}%
\pgfusepath{stroke}%
\end{pgfscope}%
\begin{pgfscope}%
\definecolor{textcolor}{rgb}{0.150000,0.150000,0.150000}%
\pgfsetstrokecolor{textcolor}%
\pgfsetfillcolor{textcolor}%
\pgftext[x=0.742901in,y=0.285273in,,top]{\color{textcolor}\rmfamily\fontsize{10.000000}{12.000000}\selectfont \(\displaystyle {\ensuremath{-}3}\)}%
\end{pgfscope}%
\begin{pgfscope}%
\pgfpathrectangle{\pgfqpoint{0.474151in}{0.333884in}}{\pgfqpoint{2.939735in}{2.337191in}}%
\pgfusepath{clip}%
\pgfsetroundcap%
\pgfsetroundjoin%
\pgfsetlinewidth{0.803000pt}%
\definecolor{currentstroke}{rgb}{0.800000,0.800000,0.800000}%
\pgfsetstrokecolor{currentstroke}%
\pgfsetdash{}{0pt}%
\pgfpathmoveto{\pgfqpoint{1.043180in}{0.333884in}}%
\pgfpathlineto{\pgfqpoint{1.043180in}{2.671076in}}%
\pgfusepath{stroke}%
\end{pgfscope}%
\begin{pgfscope}%
\definecolor{textcolor}{rgb}{0.150000,0.150000,0.150000}%
\pgfsetstrokecolor{textcolor}%
\pgfsetfillcolor{textcolor}%
\pgftext[x=1.043180in,y=0.285273in,,top]{\color{textcolor}\rmfamily\fontsize{10.000000}{12.000000}\selectfont \(\displaystyle {\ensuremath{-}2}\)}%
\end{pgfscope}%
\begin{pgfscope}%
\pgfpathrectangle{\pgfqpoint{0.474151in}{0.333884in}}{\pgfqpoint{2.939735in}{2.337191in}}%
\pgfusepath{clip}%
\pgfsetroundcap%
\pgfsetroundjoin%
\pgfsetlinewidth{0.803000pt}%
\definecolor{currentstroke}{rgb}{0.800000,0.800000,0.800000}%
\pgfsetstrokecolor{currentstroke}%
\pgfsetdash{}{0pt}%
\pgfpathmoveto{\pgfqpoint{1.343459in}{0.333884in}}%
\pgfpathlineto{\pgfqpoint{1.343459in}{2.671076in}}%
\pgfusepath{stroke}%
\end{pgfscope}%
\begin{pgfscope}%
\definecolor{textcolor}{rgb}{0.150000,0.150000,0.150000}%
\pgfsetstrokecolor{textcolor}%
\pgfsetfillcolor{textcolor}%
\pgftext[x=1.343459in,y=0.285273in,,top]{\color{textcolor}\rmfamily\fontsize{10.000000}{12.000000}\selectfont \(\displaystyle {\ensuremath{-}1}\)}%
\end{pgfscope}%
\begin{pgfscope}%
\pgfpathrectangle{\pgfqpoint{0.474151in}{0.333884in}}{\pgfqpoint{2.939735in}{2.337191in}}%
\pgfusepath{clip}%
\pgfsetroundcap%
\pgfsetroundjoin%
\pgfsetlinewidth{0.803000pt}%
\definecolor{currentstroke}{rgb}{0.800000,0.800000,0.800000}%
\pgfsetstrokecolor{currentstroke}%
\pgfsetdash{}{0pt}%
\pgfpathmoveto{\pgfqpoint{1.643739in}{0.333884in}}%
\pgfpathlineto{\pgfqpoint{1.643739in}{2.671076in}}%
\pgfusepath{stroke}%
\end{pgfscope}%
\begin{pgfscope}%
\definecolor{textcolor}{rgb}{0.150000,0.150000,0.150000}%
\pgfsetstrokecolor{textcolor}%
\pgfsetfillcolor{textcolor}%
\pgftext[x=1.643739in,y=0.285273in,,top]{\color{textcolor}\rmfamily\fontsize{10.000000}{12.000000}\selectfont \(\displaystyle {0}\)}%
\end{pgfscope}%
\begin{pgfscope}%
\pgfpathrectangle{\pgfqpoint{0.474151in}{0.333884in}}{\pgfqpoint{2.939735in}{2.337191in}}%
\pgfusepath{clip}%
\pgfsetroundcap%
\pgfsetroundjoin%
\pgfsetlinewidth{0.803000pt}%
\definecolor{currentstroke}{rgb}{0.800000,0.800000,0.800000}%
\pgfsetstrokecolor{currentstroke}%
\pgfsetdash{}{0pt}%
\pgfpathmoveto{\pgfqpoint{1.944018in}{0.333884in}}%
\pgfpathlineto{\pgfqpoint{1.944018in}{2.671076in}}%
\pgfusepath{stroke}%
\end{pgfscope}%
\begin{pgfscope}%
\definecolor{textcolor}{rgb}{0.150000,0.150000,0.150000}%
\pgfsetstrokecolor{textcolor}%
\pgfsetfillcolor{textcolor}%
\pgftext[x=1.944018in,y=0.285273in,,top]{\color{textcolor}\rmfamily\fontsize{10.000000}{12.000000}\selectfont \(\displaystyle {1}\)}%
\end{pgfscope}%
\begin{pgfscope}%
\pgfpathrectangle{\pgfqpoint{0.474151in}{0.333884in}}{\pgfqpoint{2.939735in}{2.337191in}}%
\pgfusepath{clip}%
\pgfsetroundcap%
\pgfsetroundjoin%
\pgfsetlinewidth{0.803000pt}%
\definecolor{currentstroke}{rgb}{0.800000,0.800000,0.800000}%
\pgfsetstrokecolor{currentstroke}%
\pgfsetdash{}{0pt}%
\pgfpathmoveto{\pgfqpoint{2.244297in}{0.333884in}}%
\pgfpathlineto{\pgfqpoint{2.244297in}{2.671076in}}%
\pgfusepath{stroke}%
\end{pgfscope}%
\begin{pgfscope}%
\definecolor{textcolor}{rgb}{0.150000,0.150000,0.150000}%
\pgfsetstrokecolor{textcolor}%
\pgfsetfillcolor{textcolor}%
\pgftext[x=2.244297in,y=0.285273in,,top]{\color{textcolor}\rmfamily\fontsize{10.000000}{12.000000}\selectfont \(\displaystyle {2}\)}%
\end{pgfscope}%
\begin{pgfscope}%
\pgfpathrectangle{\pgfqpoint{0.474151in}{0.333884in}}{\pgfqpoint{2.939735in}{2.337191in}}%
\pgfusepath{clip}%
\pgfsetroundcap%
\pgfsetroundjoin%
\pgfsetlinewidth{0.803000pt}%
\definecolor{currentstroke}{rgb}{0.800000,0.800000,0.800000}%
\pgfsetstrokecolor{currentstroke}%
\pgfsetdash{}{0pt}%
\pgfpathmoveto{\pgfqpoint{2.544577in}{0.333884in}}%
\pgfpathlineto{\pgfqpoint{2.544577in}{2.671076in}}%
\pgfusepath{stroke}%
\end{pgfscope}%
\begin{pgfscope}%
\definecolor{textcolor}{rgb}{0.150000,0.150000,0.150000}%
\pgfsetstrokecolor{textcolor}%
\pgfsetfillcolor{textcolor}%
\pgftext[x=2.544577in,y=0.285273in,,top]{\color{textcolor}\rmfamily\fontsize{10.000000}{12.000000}\selectfont \(\displaystyle {3}\)}%
\end{pgfscope}%
\begin{pgfscope}%
\pgfpathrectangle{\pgfqpoint{0.474151in}{0.333884in}}{\pgfqpoint{2.939735in}{2.337191in}}%
\pgfusepath{clip}%
\pgfsetroundcap%
\pgfsetroundjoin%
\pgfsetlinewidth{0.803000pt}%
\definecolor{currentstroke}{rgb}{0.800000,0.800000,0.800000}%
\pgfsetstrokecolor{currentstroke}%
\pgfsetdash{}{0pt}%
\pgfpathmoveto{\pgfqpoint{2.844856in}{0.333884in}}%
\pgfpathlineto{\pgfqpoint{2.844856in}{2.671076in}}%
\pgfusepath{stroke}%
\end{pgfscope}%
\begin{pgfscope}%
\definecolor{textcolor}{rgb}{0.150000,0.150000,0.150000}%
\pgfsetstrokecolor{textcolor}%
\pgfsetfillcolor{textcolor}%
\pgftext[x=2.844856in,y=0.285273in,,top]{\color{textcolor}\rmfamily\fontsize{10.000000}{12.000000}\selectfont \(\displaystyle {4}\)}%
\end{pgfscope}%
\begin{pgfscope}%
\pgfpathrectangle{\pgfqpoint{0.474151in}{0.333884in}}{\pgfqpoint{2.939735in}{2.337191in}}%
\pgfusepath{clip}%
\pgfsetroundcap%
\pgfsetroundjoin%
\pgfsetlinewidth{0.803000pt}%
\definecolor{currentstroke}{rgb}{0.800000,0.800000,0.800000}%
\pgfsetstrokecolor{currentstroke}%
\pgfsetdash{}{0pt}%
\pgfpathmoveto{\pgfqpoint{3.145135in}{0.333884in}}%
\pgfpathlineto{\pgfqpoint{3.145135in}{2.671076in}}%
\pgfusepath{stroke}%
\end{pgfscope}%
\begin{pgfscope}%
\definecolor{textcolor}{rgb}{0.150000,0.150000,0.150000}%
\pgfsetstrokecolor{textcolor}%
\pgfsetfillcolor{textcolor}%
\pgftext[x=3.145135in,y=0.285273in,,top]{\color{textcolor}\rmfamily\fontsize{10.000000}{12.000000}\selectfont \(\displaystyle {5}\)}%
\end{pgfscope}%
\begin{pgfscope}%
\definecolor{textcolor}{rgb}{0.150000,0.150000,0.150000}%
\pgfsetstrokecolor{textcolor}%
\pgfsetfillcolor{textcolor}%
\pgftext[x=1.944018in,y=0.106385in,,top]{\color{textcolor}\rmfamily\fontsize{10.000000}{12.000000}\selectfont Goals difference}%
\end{pgfscope}%
\begin{pgfscope}%
\pgfpathrectangle{\pgfqpoint{0.474151in}{0.333884in}}{\pgfqpoint{2.939735in}{2.337191in}}%
\pgfusepath{clip}%
\pgfsetroundcap%
\pgfsetroundjoin%
\pgfsetlinewidth{0.803000pt}%
\definecolor{currentstroke}{rgb}{0.800000,0.800000,0.800000}%
\pgfsetstrokecolor{currentstroke}%
\pgfsetdash{}{0pt}%
\pgfpathmoveto{\pgfqpoint{0.474151in}{0.333884in}}%
\pgfpathlineto{\pgfqpoint{3.413885in}{0.333884in}}%
\pgfusepath{stroke}%
\end{pgfscope}%
\begin{pgfscope}%
\definecolor{textcolor}{rgb}{0.150000,0.150000,0.150000}%
\pgfsetstrokecolor{textcolor}%
\pgfsetfillcolor{textcolor}%
\pgftext[x=0.248070in, y=0.285690in, left, base]{\color{textcolor}\rmfamily\fontsize{10.000000}{12.000000}\selectfont \(\displaystyle {0.0}\)}%
\end{pgfscope}%
\begin{pgfscope}%
\pgfpathrectangle{\pgfqpoint{0.474151in}{0.333884in}}{\pgfqpoint{2.939735in}{2.337191in}}%
\pgfusepath{clip}%
\pgfsetroundcap%
\pgfsetroundjoin%
\pgfsetlinewidth{0.803000pt}%
\definecolor{currentstroke}{rgb}{0.800000,0.800000,0.800000}%
\pgfsetstrokecolor{currentstroke}%
\pgfsetdash{}{0pt}%
\pgfpathmoveto{\pgfqpoint{0.474151in}{0.853260in}}%
\pgfpathlineto{\pgfqpoint{3.413885in}{0.853260in}}%
\pgfusepath{stroke}%
\end{pgfscope}%
\begin{pgfscope}%
\definecolor{textcolor}{rgb}{0.150000,0.150000,0.150000}%
\pgfsetstrokecolor{textcolor}%
\pgfsetfillcolor{textcolor}%
\pgftext[x=0.248070in, y=0.805066in, left, base]{\color{textcolor}\rmfamily\fontsize{10.000000}{12.000000}\selectfont \(\displaystyle {0.1}\)}%
\end{pgfscope}%
\begin{pgfscope}%
\pgfpathrectangle{\pgfqpoint{0.474151in}{0.333884in}}{\pgfqpoint{2.939735in}{2.337191in}}%
\pgfusepath{clip}%
\pgfsetroundcap%
\pgfsetroundjoin%
\pgfsetlinewidth{0.803000pt}%
\definecolor{currentstroke}{rgb}{0.800000,0.800000,0.800000}%
\pgfsetstrokecolor{currentstroke}%
\pgfsetdash{}{0pt}%
\pgfpathmoveto{\pgfqpoint{0.474151in}{1.372636in}}%
\pgfpathlineto{\pgfqpoint{3.413885in}{1.372636in}}%
\pgfusepath{stroke}%
\end{pgfscope}%
\begin{pgfscope}%
\definecolor{textcolor}{rgb}{0.150000,0.150000,0.150000}%
\pgfsetstrokecolor{textcolor}%
\pgfsetfillcolor{textcolor}%
\pgftext[x=0.248070in, y=1.324442in, left, base]{\color{textcolor}\rmfamily\fontsize{10.000000}{12.000000}\selectfont \(\displaystyle {0.2}\)}%
\end{pgfscope}%
\begin{pgfscope}%
\pgfpathrectangle{\pgfqpoint{0.474151in}{0.333884in}}{\pgfqpoint{2.939735in}{2.337191in}}%
\pgfusepath{clip}%
\pgfsetroundcap%
\pgfsetroundjoin%
\pgfsetlinewidth{0.803000pt}%
\definecolor{currentstroke}{rgb}{0.800000,0.800000,0.800000}%
\pgfsetstrokecolor{currentstroke}%
\pgfsetdash{}{0pt}%
\pgfpathmoveto{\pgfqpoint{0.474151in}{1.892012in}}%
\pgfpathlineto{\pgfqpoint{3.413885in}{1.892012in}}%
\pgfusepath{stroke}%
\end{pgfscope}%
\begin{pgfscope}%
\definecolor{textcolor}{rgb}{0.150000,0.150000,0.150000}%
\pgfsetstrokecolor{textcolor}%
\pgfsetfillcolor{textcolor}%
\pgftext[x=0.248070in, y=1.843817in, left, base]{\color{textcolor}\rmfamily\fontsize{10.000000}{12.000000}\selectfont \(\displaystyle {0.3}\)}%
\end{pgfscope}%
\begin{pgfscope}%
\pgfpathrectangle{\pgfqpoint{0.474151in}{0.333884in}}{\pgfqpoint{2.939735in}{2.337191in}}%
\pgfusepath{clip}%
\pgfsetroundcap%
\pgfsetroundjoin%
\pgfsetlinewidth{0.803000pt}%
\definecolor{currentstroke}{rgb}{0.800000,0.800000,0.800000}%
\pgfsetstrokecolor{currentstroke}%
\pgfsetdash{}{0pt}%
\pgfpathmoveto{\pgfqpoint{0.474151in}{2.411388in}}%
\pgfpathlineto{\pgfqpoint{3.413885in}{2.411388in}}%
\pgfusepath{stroke}%
\end{pgfscope}%
\begin{pgfscope}%
\definecolor{textcolor}{rgb}{0.150000,0.150000,0.150000}%
\pgfsetstrokecolor{textcolor}%
\pgfsetfillcolor{textcolor}%
\pgftext[x=0.248070in, y=2.363193in, left, base]{\color{textcolor}\rmfamily\fontsize{10.000000}{12.000000}\selectfont \(\displaystyle {0.4}\)}%
\end{pgfscope}%
\begin{pgfscope}%
\pgfpathrectangle{\pgfqpoint{0.474151in}{0.333884in}}{\pgfqpoint{2.939735in}{2.337191in}}%
\pgfusepath{clip}%
\pgfsetroundcap%
\pgfsetroundjoin%
\pgfsetlinewidth{0.803000pt}%
\definecolor{currentstroke}{rgb}{0.800000,0.800000,0.800000}%
\pgfsetstrokecolor{currentstroke}%
\pgfsetdash{}{0pt}%
\pgfpathmoveto{\pgfqpoint{0.474151in}{0.593572in}}%
\pgfpathlineto{\pgfqpoint{3.413885in}{0.593572in}}%
\pgfusepath{stroke}%
\end{pgfscope}%
\begin{pgfscope}%
\pgfpathrectangle{\pgfqpoint{0.474151in}{0.333884in}}{\pgfqpoint{2.939735in}{2.337191in}}%
\pgfusepath{clip}%
\pgfsetroundcap%
\pgfsetroundjoin%
\pgfsetlinewidth{0.803000pt}%
\definecolor{currentstroke}{rgb}{0.800000,0.800000,0.800000}%
\pgfsetstrokecolor{currentstroke}%
\pgfsetdash{}{0pt}%
\pgfpathmoveto{\pgfqpoint{0.474151in}{1.112948in}}%
\pgfpathlineto{\pgfqpoint{3.413885in}{1.112948in}}%
\pgfusepath{stroke}%
\end{pgfscope}%
\begin{pgfscope}%
\pgfpathrectangle{\pgfqpoint{0.474151in}{0.333884in}}{\pgfqpoint{2.939735in}{2.337191in}}%
\pgfusepath{clip}%
\pgfsetroundcap%
\pgfsetroundjoin%
\pgfsetlinewidth{0.803000pt}%
\definecolor{currentstroke}{rgb}{0.800000,0.800000,0.800000}%
\pgfsetstrokecolor{currentstroke}%
\pgfsetdash{}{0pt}%
\pgfpathmoveto{\pgfqpoint{0.474151in}{1.632324in}}%
\pgfpathlineto{\pgfqpoint{3.413885in}{1.632324in}}%
\pgfusepath{stroke}%
\end{pgfscope}%
\begin{pgfscope}%
\pgfpathrectangle{\pgfqpoint{0.474151in}{0.333884in}}{\pgfqpoint{2.939735in}{2.337191in}}%
\pgfusepath{clip}%
\pgfsetroundcap%
\pgfsetroundjoin%
\pgfsetlinewidth{0.803000pt}%
\definecolor{currentstroke}{rgb}{0.800000,0.800000,0.800000}%
\pgfsetstrokecolor{currentstroke}%
\pgfsetdash{}{0pt}%
\pgfpathmoveto{\pgfqpoint{0.474151in}{2.151700in}}%
\pgfpathlineto{\pgfqpoint{3.413885in}{2.151700in}}%
\pgfusepath{stroke}%
\end{pgfscope}%
\begin{pgfscope}%
\pgfpathrectangle{\pgfqpoint{0.474151in}{0.333884in}}{\pgfqpoint{2.939735in}{2.337191in}}%
\pgfusepath{clip}%
\pgfsetroundcap%
\pgfsetroundjoin%
\pgfsetlinewidth{0.803000pt}%
\definecolor{currentstroke}{rgb}{0.800000,0.800000,0.800000}%
\pgfsetstrokecolor{currentstroke}%
\pgfsetdash{}{0pt}%
\pgfpathmoveto{\pgfqpoint{0.474151in}{2.671076in}}%
\pgfpathlineto{\pgfqpoint{3.413885in}{2.671076in}}%
\pgfusepath{stroke}%
\end{pgfscope}%
\begin{pgfscope}%
\definecolor{textcolor}{rgb}{0.150000,0.150000,0.150000}%
\pgfsetstrokecolor{textcolor}%
\pgfsetfillcolor{textcolor}%
\pgftext[x=0.192514in,y=1.502480in,,bottom,rotate=90.000000]{\color{textcolor}\rmfamily\fontsize{10.000000}{12.000000}\selectfont Probability density}%
\end{pgfscope}%
\begin{pgfscope}%
\pgfpathrectangle{\pgfqpoint{0.474151in}{0.333884in}}{\pgfqpoint{2.939735in}{2.337191in}}%
\pgfusepath{clip}%
\pgfsetbuttcap%
\pgfsetmiterjoin%
\definecolor{currentfill}{rgb}{0.121569,0.466667,0.705882}%
\pgfsetfillcolor{currentfill}%
\pgfsetlinewidth{0.000000pt}%
\definecolor{currentstroke}{rgb}{0.000000,0.000000,0.000000}%
\pgfsetstrokecolor{currentstroke}%
\pgfsetstrokeopacity{0.000000}%
\pgfsetdash{}{0pt}%
\pgfpathmoveto{\pgfqpoint{0.607775in}{0.333884in}}%
\pgfpathlineto{\pgfqpoint{0.878026in}{0.333884in}}%
\pgfpathlineto{\pgfqpoint{0.878026in}{0.396210in}}%
\pgfpathlineto{\pgfqpoint{0.607775in}{0.396210in}}%
\pgfpathlineto{\pgfqpoint{0.607775in}{0.333884in}}%
\pgfpathclose%
\pgfusepath{fill}%
\end{pgfscope}%
\begin{pgfscope}%
\pgfpathrectangle{\pgfqpoint{0.474151in}{0.333884in}}{\pgfqpoint{2.939735in}{2.337191in}}%
\pgfusepath{clip}%
\pgfsetbuttcap%
\pgfsetmiterjoin%
\definecolor{currentfill}{rgb}{0.121569,0.466667,0.705882}%
\pgfsetfillcolor{currentfill}%
\pgfsetlinewidth{0.000000pt}%
\definecolor{currentstroke}{rgb}{0.000000,0.000000,0.000000}%
\pgfsetstrokecolor{currentstroke}%
\pgfsetstrokeopacity{0.000000}%
\pgfsetdash{}{0pt}%
\pgfpathmoveto{\pgfqpoint{0.908054in}{0.333884in}}%
\pgfpathlineto{\pgfqpoint{1.178306in}{0.333884in}}%
\pgfpathlineto{\pgfqpoint{1.178306in}{0.583185in}}%
\pgfpathlineto{\pgfqpoint{0.908054in}{0.583185in}}%
\pgfpathlineto{\pgfqpoint{0.908054in}{0.333884in}}%
\pgfpathclose%
\pgfusepath{fill}%
\end{pgfscope}%
\begin{pgfscope}%
\pgfpathrectangle{\pgfqpoint{0.474151in}{0.333884in}}{\pgfqpoint{2.939735in}{2.337191in}}%
\pgfusepath{clip}%
\pgfsetbuttcap%
\pgfsetmiterjoin%
\definecolor{currentfill}{rgb}{0.121569,0.466667,0.705882}%
\pgfsetfillcolor{currentfill}%
\pgfsetlinewidth{0.000000pt}%
\definecolor{currentstroke}{rgb}{0.000000,0.000000,0.000000}%
\pgfsetstrokecolor{currentstroke}%
\pgfsetstrokeopacity{0.000000}%
\pgfsetdash{}{0pt}%
\pgfpathmoveto{\pgfqpoint{1.208334in}{0.333884in}}%
\pgfpathlineto{\pgfqpoint{1.478585in}{0.333884in}}%
\pgfpathlineto{\pgfqpoint{1.478585in}{1.206436in}}%
\pgfpathlineto{\pgfqpoint{1.208334in}{1.206436in}}%
\pgfpathlineto{\pgfqpoint{1.208334in}{0.333884in}}%
\pgfpathclose%
\pgfusepath{fill}%
\end{pgfscope}%
\begin{pgfscope}%
\pgfpathrectangle{\pgfqpoint{0.474151in}{0.333884in}}{\pgfqpoint{2.939735in}{2.337191in}}%
\pgfusepath{clip}%
\pgfsetbuttcap%
\pgfsetmiterjoin%
\definecolor{currentfill}{rgb}{0.121569,0.466667,0.705882}%
\pgfsetfillcolor{currentfill}%
\pgfsetlinewidth{0.000000pt}%
\definecolor{currentstroke}{rgb}{0.000000,0.000000,0.000000}%
\pgfsetstrokecolor{currentstroke}%
\pgfsetstrokeopacity{0.000000}%
\pgfsetdash{}{0pt}%
\pgfpathmoveto{\pgfqpoint{1.508613in}{0.333884in}}%
\pgfpathlineto{\pgfqpoint{1.778864in}{0.333884in}}%
\pgfpathlineto{\pgfqpoint{1.778864in}{2.536038in}}%
\pgfpathlineto{\pgfqpoint{1.508613in}{2.536038in}}%
\pgfpathlineto{\pgfqpoint{1.508613in}{0.333884in}}%
\pgfpathclose%
\pgfusepath{fill}%
\end{pgfscope}%
\begin{pgfscope}%
\pgfpathrectangle{\pgfqpoint{0.474151in}{0.333884in}}{\pgfqpoint{2.939735in}{2.337191in}}%
\pgfusepath{clip}%
\pgfsetbuttcap%
\pgfsetmiterjoin%
\definecolor{currentfill}{rgb}{0.121569,0.466667,0.705882}%
\pgfsetfillcolor{currentfill}%
\pgfsetlinewidth{0.000000pt}%
\definecolor{currentstroke}{rgb}{0.000000,0.000000,0.000000}%
\pgfsetstrokecolor{currentstroke}%
\pgfsetstrokeopacity{0.000000}%
\pgfsetdash{}{0pt}%
\pgfpathmoveto{\pgfqpoint{1.808892in}{0.333884in}}%
\pgfpathlineto{\pgfqpoint{2.079144in}{0.333884in}}%
\pgfpathlineto{\pgfqpoint{2.079144in}{1.538836in}}%
\pgfpathlineto{\pgfqpoint{1.808892in}{1.538836in}}%
\pgfpathlineto{\pgfqpoint{1.808892in}{0.333884in}}%
\pgfpathclose%
\pgfusepath{fill}%
\end{pgfscope}%
\begin{pgfscope}%
\pgfpathrectangle{\pgfqpoint{0.474151in}{0.333884in}}{\pgfqpoint{2.939735in}{2.337191in}}%
\pgfusepath{clip}%
\pgfsetbuttcap%
\pgfsetmiterjoin%
\definecolor{currentfill}{rgb}{0.121569,0.466667,0.705882}%
\pgfsetfillcolor{currentfill}%
\pgfsetlinewidth{0.000000pt}%
\definecolor{currentstroke}{rgb}{0.000000,0.000000,0.000000}%
\pgfsetstrokecolor{currentstroke}%
\pgfsetstrokeopacity{0.000000}%
\pgfsetdash{}{0pt}%
\pgfpathmoveto{\pgfqpoint{2.109172in}{0.333884in}}%
\pgfpathlineto{\pgfqpoint{2.379423in}{0.333884in}}%
\pgfpathlineto{\pgfqpoint{2.379423in}{0.749385in}}%
\pgfpathlineto{\pgfqpoint{2.109172in}{0.749385in}}%
\pgfpathlineto{\pgfqpoint{2.109172in}{0.333884in}}%
\pgfpathclose%
\pgfusepath{fill}%
\end{pgfscope}%
\begin{pgfscope}%
\pgfpathrectangle{\pgfqpoint{0.474151in}{0.333884in}}{\pgfqpoint{2.939735in}{2.337191in}}%
\pgfusepath{clip}%
\pgfsetbuttcap%
\pgfsetmiterjoin%
\definecolor{currentfill}{rgb}{0.121569,0.466667,0.705882}%
\pgfsetfillcolor{currentfill}%
\pgfsetlinewidth{0.000000pt}%
\definecolor{currentstroke}{rgb}{0.000000,0.000000,0.000000}%
\pgfsetstrokecolor{currentstroke}%
\pgfsetstrokeopacity{0.000000}%
\pgfsetdash{}{0pt}%
\pgfpathmoveto{\pgfqpoint{2.409451in}{0.333884in}}%
\pgfpathlineto{\pgfqpoint{2.679702in}{0.333884in}}%
\pgfpathlineto{\pgfqpoint{2.679702in}{0.479310in}}%
\pgfpathlineto{\pgfqpoint{2.409451in}{0.479310in}}%
\pgfpathlineto{\pgfqpoint{2.409451in}{0.333884in}}%
\pgfpathclose%
\pgfusepath{fill}%
\end{pgfscope}%
\begin{pgfscope}%
\pgfpathrectangle{\pgfqpoint{0.474151in}{0.333884in}}{\pgfqpoint{2.939735in}{2.337191in}}%
\pgfusepath{clip}%
\pgfsetbuttcap%
\pgfsetmiterjoin%
\definecolor{currentfill}{rgb}{0.121569,0.466667,0.705882}%
\pgfsetfillcolor{currentfill}%
\pgfsetlinewidth{0.000000pt}%
\definecolor{currentstroke}{rgb}{0.000000,0.000000,0.000000}%
\pgfsetstrokecolor{currentstroke}%
\pgfsetstrokeopacity{0.000000}%
\pgfsetdash{}{0pt}%
\pgfpathmoveto{\pgfqpoint{2.709730in}{0.333884in}}%
\pgfpathlineto{\pgfqpoint{2.979982in}{0.333884in}}%
\pgfpathlineto{\pgfqpoint{2.979982in}{0.354659in}}%
\pgfpathlineto{\pgfqpoint{2.709730in}{0.354659in}}%
\pgfpathlineto{\pgfqpoint{2.709730in}{0.333884in}}%
\pgfpathclose%
\pgfusepath{fill}%
\end{pgfscope}%
\begin{pgfscope}%
\pgfpathrectangle{\pgfqpoint{0.474151in}{0.333884in}}{\pgfqpoint{2.939735in}{2.337191in}}%
\pgfusepath{clip}%
\pgfsetbuttcap%
\pgfsetmiterjoin%
\definecolor{currentfill}{rgb}{0.121569,0.466667,0.705882}%
\pgfsetfillcolor{currentfill}%
\pgfsetlinewidth{0.000000pt}%
\definecolor{currentstroke}{rgb}{0.000000,0.000000,0.000000}%
\pgfsetstrokecolor{currentstroke}%
\pgfsetstrokeopacity{0.000000}%
\pgfsetdash{}{0pt}%
\pgfpathmoveto{\pgfqpoint{3.010010in}{0.333884in}}%
\pgfpathlineto{\pgfqpoint{3.280261in}{0.333884in}}%
\pgfpathlineto{\pgfqpoint{3.280261in}{0.354659in}}%
\pgfpathlineto{\pgfqpoint{3.010010in}{0.354659in}}%
\pgfpathlineto{\pgfqpoint{3.010010in}{0.333884in}}%
\pgfpathclose%
\pgfusepath{fill}%
\end{pgfscope}%
\begin{pgfscope}%
\pgfpathrectangle{\pgfqpoint{0.474151in}{0.333884in}}{\pgfqpoint{2.939735in}{2.337191in}}%
\pgfusepath{clip}%
\pgfsetbuttcap%
\pgfsetroundjoin%
\pgfsetlinewidth{3.011250pt}%
\definecolor{currentstroke}{rgb}{1.000000,0.498039,0.054902}%
\pgfsetstrokecolor{currentstroke}%
\pgfsetdash{{11.100000pt}{4.800000pt}}{0.000000pt}%
\pgfpathmoveto{\pgfqpoint{0.742901in}{0.376959in}}%
\pgfpathlineto{\pgfqpoint{0.767166in}{0.385673in}}%
\pgfpathlineto{\pgfqpoint{0.791431in}{0.395859in}}%
\pgfpathlineto{\pgfqpoint{0.815696in}{0.407700in}}%
\pgfpathlineto{\pgfqpoint{0.839961in}{0.421392in}}%
\pgfpathlineto{\pgfqpoint{0.864226in}{0.437139in}}%
\pgfpathlineto{\pgfqpoint{0.888491in}{0.455148in}}%
\pgfpathlineto{\pgfqpoint{0.912756in}{0.475631in}}%
\pgfpathlineto{\pgfqpoint{0.937021in}{0.498798in}}%
\pgfpathlineto{\pgfqpoint{0.961286in}{0.524854in}}%
\pgfpathlineto{\pgfqpoint{0.985551in}{0.553990in}}%
\pgfpathlineto{\pgfqpoint{1.009816in}{0.586383in}}%
\pgfpathlineto{\pgfqpoint{1.034081in}{0.622187in}}%
\pgfpathlineto{\pgfqpoint{1.058346in}{0.661526in}}%
\pgfpathlineto{\pgfqpoint{1.082611in}{0.704490in}}%
\pgfpathlineto{\pgfqpoint{1.106876in}{0.751124in}}%
\pgfpathlineto{\pgfqpoint{1.131141in}{0.801426in}}%
\pgfpathlineto{\pgfqpoint{1.155406in}{0.855339in}}%
\pgfpathlineto{\pgfqpoint{1.179671in}{0.912745in}}%
\pgfpathlineto{\pgfqpoint{1.203936in}{0.973462in}}%
\pgfpathlineto{\pgfqpoint{1.228201in}{1.037238in}}%
\pgfpathlineto{\pgfqpoint{1.252466in}{1.103750in}}%
\pgfpathlineto{\pgfqpoint{1.276731in}{1.172606in}}%
\pgfpathlineto{\pgfqpoint{1.300996in}{1.243341in}}%
\pgfpathlineto{\pgfqpoint{1.325261in}{1.315423in}}%
\pgfpathlineto{\pgfqpoint{1.349526in}{1.388258in}}%
\pgfpathlineto{\pgfqpoint{1.373791in}{1.461192in}}%
\pgfpathlineto{\pgfqpoint{1.398056in}{1.533527in}}%
\pgfpathlineto{\pgfqpoint{1.422321in}{1.604525in}}%
\pgfpathlineto{\pgfqpoint{1.446586in}{1.673422in}}%
\pgfpathlineto{\pgfqpoint{1.470851in}{1.739441in}}%
\pgfpathlineto{\pgfqpoint{1.495116in}{1.801806in}}%
\pgfpathlineto{\pgfqpoint{1.519381in}{1.859760in}}%
\pgfpathlineto{\pgfqpoint{1.543646in}{1.912573in}}%
\pgfpathlineto{\pgfqpoint{1.567911in}{1.959565in}}%
\pgfpathlineto{\pgfqpoint{1.592176in}{2.000115in}}%
\pgfpathlineto{\pgfqpoint{1.616441in}{2.033679in}}%
\pgfpathlineto{\pgfqpoint{1.640706in}{2.059798in}}%
\pgfpathlineto{\pgfqpoint{1.664971in}{2.078112in}}%
\pgfpathlineto{\pgfqpoint{1.689236in}{2.088364in}}%
\pgfpathlineto{\pgfqpoint{1.713501in}{2.090412in}}%
\pgfpathlineto{\pgfqpoint{1.737766in}{2.084227in}}%
\pgfpathlineto{\pgfqpoint{1.762031in}{2.069894in}}%
\pgfpathlineto{\pgfqpoint{1.786296in}{2.047616in}}%
\pgfpathlineto{\pgfqpoint{1.810561in}{2.017701in}}%
\pgfpathlineto{\pgfqpoint{1.834826in}{1.980559in}}%
\pgfpathlineto{\pgfqpoint{1.859091in}{1.936696in}}%
\pgfpathlineto{\pgfqpoint{1.883356in}{1.886694in}}%
\pgfpathlineto{\pgfqpoint{1.907621in}{1.831207in}}%
\pgfpathlineto{\pgfqpoint{1.931886in}{1.770941in}}%
\pgfpathlineto{\pgfqpoint{1.956151in}{1.706641in}}%
\pgfpathlineto{\pgfqpoint{1.980416in}{1.639078in}}%
\pgfpathlineto{\pgfqpoint{2.004681in}{1.569027in}}%
\pgfpathlineto{\pgfqpoint{2.028946in}{1.497263in}}%
\pgfpathlineto{\pgfqpoint{2.053211in}{1.424536in}}%
\pgfpathlineto{\pgfqpoint{2.077476in}{1.351567in}}%
\pgfpathlineto{\pgfqpoint{2.101741in}{1.279033in}}%
\pgfpathlineto{\pgfqpoint{2.126006in}{1.207558in}}%
\pgfpathlineto{\pgfqpoint{2.150271in}{1.137706in}}%
\pgfpathlineto{\pgfqpoint{2.174536in}{1.069975in}}%
\pgfpathlineto{\pgfqpoint{2.198801in}{1.004794in}}%
\pgfpathlineto{\pgfqpoint{2.223066in}{0.942522in}}%
\pgfpathlineto{\pgfqpoint{2.247331in}{0.883443in}}%
\pgfpathlineto{\pgfqpoint{2.271596in}{0.827775in}}%
\pgfpathlineto{\pgfqpoint{2.295861in}{0.775668in}}%
\pgfpathlineto{\pgfqpoint{2.320126in}{0.727207in}}%
\pgfpathlineto{\pgfqpoint{2.344391in}{0.682422in}}%
\pgfpathlineto{\pgfqpoint{2.368656in}{0.641290in}}%
\pgfpathlineto{\pgfqpoint{2.392921in}{0.603743in}}%
\pgfpathlineto{\pgfqpoint{2.417186in}{0.569672in}}%
\pgfpathlineto{\pgfqpoint{2.441451in}{0.538938in}}%
\pgfpathlineto{\pgfqpoint{2.465716in}{0.511375in}}%
\pgfpathlineto{\pgfqpoint{2.489981in}{0.486797in}}%
\pgfpathlineto{\pgfqpoint{2.514246in}{0.465006in}}%
\pgfpathlineto{\pgfqpoint{2.538511in}{0.445793in}}%
\pgfpathlineto{\pgfqpoint{2.562776in}{0.428949in}}%
\pgfpathlineto{\pgfqpoint{2.587041in}{0.414261in}}%
\pgfpathlineto{\pgfqpoint{2.611305in}{0.401525in}}%
\pgfpathlineto{\pgfqpoint{2.635570in}{0.390540in}}%
\pgfpathlineto{\pgfqpoint{2.659835in}{0.381117in}}%
\pgfpathlineto{\pgfqpoint{2.684100in}{0.373077in}}%
\pgfpathlineto{\pgfqpoint{2.708365in}{0.366253in}}%
\pgfpathlineto{\pgfqpoint{2.732630in}{0.360492in}}%
\pgfpathlineto{\pgfqpoint{2.756895in}{0.355654in}}%
\pgfpathlineto{\pgfqpoint{2.781160in}{0.351612in}}%
\pgfpathlineto{\pgfqpoint{2.805425in}{0.348253in}}%
\pgfpathlineto{\pgfqpoint{2.829690in}{0.345476in}}%
\pgfpathlineto{\pgfqpoint{2.853955in}{0.343192in}}%
\pgfpathlineto{\pgfqpoint{2.878220in}{0.341323in}}%
\pgfpathlineto{\pgfqpoint{2.902485in}{0.339801in}}%
\pgfpathlineto{\pgfqpoint{2.926750in}{0.338569in}}%
\pgfpathlineto{\pgfqpoint{2.951015in}{0.337576in}}%
\pgfpathlineto{\pgfqpoint{2.975280in}{0.336780in}}%
\pgfpathlineto{\pgfqpoint{2.999545in}{0.336145in}}%
\pgfpathlineto{\pgfqpoint{3.023810in}{0.335641in}}%
\pgfpathlineto{\pgfqpoint{3.048075in}{0.335243in}}%
\pgfpathlineto{\pgfqpoint{3.072340in}{0.334930in}}%
\pgfpathlineto{\pgfqpoint{3.096605in}{0.334685in}}%
\pgfpathlineto{\pgfqpoint{3.120870in}{0.334495in}}%
\pgfpathlineto{\pgfqpoint{3.145135in}{0.334348in}}%
\pgfusepath{stroke}%
\end{pgfscope}%
\begin{pgfscope}%
\pgfsetrectcap%
\pgfsetmiterjoin%
\pgfsetlinewidth{1.003750pt}%
\definecolor{currentstroke}{rgb}{0.800000,0.800000,0.800000}%
\pgfsetstrokecolor{currentstroke}%
\pgfsetdash{}{0pt}%
\pgfpathmoveto{\pgfqpoint{0.474151in}{0.333884in}}%
\pgfpathlineto{\pgfqpoint{0.474151in}{2.671076in}}%
\pgfusepath{stroke}%
\end{pgfscope}%
\begin{pgfscope}%
\pgfsetrectcap%
\pgfsetmiterjoin%
\pgfsetlinewidth{1.003750pt}%
\definecolor{currentstroke}{rgb}{0.800000,0.800000,0.800000}%
\pgfsetstrokecolor{currentstroke}%
\pgfsetdash{}{0pt}%
\pgfpathmoveto{\pgfqpoint{3.413885in}{0.333884in}}%
\pgfpathlineto{\pgfqpoint{3.413885in}{2.671076in}}%
\pgfusepath{stroke}%
\end{pgfscope}%
\begin{pgfscope}%
\pgfsetrectcap%
\pgfsetmiterjoin%
\pgfsetlinewidth{1.003750pt}%
\definecolor{currentstroke}{rgb}{0.800000,0.800000,0.800000}%
\pgfsetstrokecolor{currentstroke}%
\pgfsetdash{}{0pt}%
\pgfpathmoveto{\pgfqpoint{0.474151in}{0.333884in}}%
\pgfpathlineto{\pgfqpoint{3.413885in}{0.333884in}}%
\pgfusepath{stroke}%
\end{pgfscope}%
\begin{pgfscope}%
\pgfsetrectcap%
\pgfsetmiterjoin%
\pgfsetlinewidth{1.003750pt}%
\definecolor{currentstroke}{rgb}{0.800000,0.800000,0.800000}%
\pgfsetstrokecolor{currentstroke}%
\pgfsetdash{}{0pt}%
\pgfpathmoveto{\pgfqpoint{0.474151in}{2.671076in}}%
\pgfpathlineto{\pgfqpoint{3.413885in}{2.671076in}}%
\pgfusepath{stroke}%
\end{pgfscope}%
\begin{pgfscope}%
\definecolor{textcolor}{rgb}{0.150000,0.150000,0.150000}%
\pgfsetstrokecolor{textcolor}%
\pgfsetfillcolor{textcolor}%
\pgftext[x=0.742901in,y=0.437876in,,bottom]{\color{textcolor}\rmfamily\fontsize{10.000000}{12.000000}\selectfont 0.012}%
\end{pgfscope}%
\begin{pgfscope}%
\definecolor{textcolor}{rgb}{0.150000,0.150000,0.150000}%
\pgfsetstrokecolor{textcolor}%
\pgfsetfillcolor{textcolor}%
\pgftext[x=1.043180in,y=0.624852in,,bottom]{\color{textcolor}\rmfamily\fontsize{10.000000}{12.000000}\selectfont 0.048}%
\end{pgfscope}%
\begin{pgfscope}%
\definecolor{textcolor}{rgb}{0.150000,0.150000,0.150000}%
\pgfsetstrokecolor{textcolor}%
\pgfsetfillcolor{textcolor}%
\pgftext[x=1.343459in,y=1.248103in,,bottom]{\color{textcolor}\rmfamily\fontsize{10.000000}{12.000000}\selectfont 0.168}%
\end{pgfscope}%
\begin{pgfscope}%
\definecolor{textcolor}{rgb}{0.150000,0.150000,0.150000}%
\pgfsetstrokecolor{textcolor}%
\pgfsetfillcolor{textcolor}%
\pgftext[x=1.643739in,y=2.577705in,,bottom]{\color{textcolor}\rmfamily\fontsize{10.000000}{12.000000}\selectfont 0.424}%
\end{pgfscope}%
\begin{pgfscope}%
\definecolor{textcolor}{rgb}{0.150000,0.150000,0.150000}%
\pgfsetstrokecolor{textcolor}%
\pgfsetfillcolor{textcolor}%
\pgftext[x=1.944018in,y=1.580503in,,bottom]{\color{textcolor}\rmfamily\fontsize{10.000000}{12.000000}\selectfont 0.232}%
\end{pgfscope}%
\begin{pgfscope}%
\definecolor{textcolor}{rgb}{0.150000,0.150000,0.150000}%
\pgfsetstrokecolor{textcolor}%
\pgfsetfillcolor{textcolor}%
\pgftext[x=2.244297in,y=0.791052in,,bottom]{\color{textcolor}\rmfamily\fontsize{10.000000}{12.000000}\selectfont 0.08}%
\end{pgfscope}%
\begin{pgfscope}%
\definecolor{textcolor}{rgb}{0.150000,0.150000,0.150000}%
\pgfsetstrokecolor{textcolor}%
\pgfsetfillcolor{textcolor}%
\pgftext[x=2.544577in,y=0.520976in,,bottom]{\color{textcolor}\rmfamily\fontsize{10.000000}{12.000000}\selectfont 0.028}%
\end{pgfscope}%
\begin{pgfscope}%
\definecolor{textcolor}{rgb}{0.150000,0.150000,0.150000}%
\pgfsetstrokecolor{textcolor}%
\pgfsetfillcolor{textcolor}%
\pgftext[x=2.844856in,y=0.396326in,,bottom]{\color{textcolor}\rmfamily\fontsize{10.000000}{12.000000}\selectfont 0.004}%
\end{pgfscope}%
\begin{pgfscope}%
\definecolor{textcolor}{rgb}{0.150000,0.150000,0.150000}%
\pgfsetstrokecolor{textcolor}%
\pgfsetfillcolor{textcolor}%
\pgftext[x=3.145135in,y=0.396326in,,bottom]{\color{textcolor}\rmfamily\fontsize{10.000000}{12.000000}\selectfont 0.004}%
\end{pgfscope}%
\end{pgfpicture}%
\makeatother%
\endgroup%

%% file: contents/conclusion.tex
\section{Conclusion}

The application case showed how it is possible to coordinate a multi-agent system with dynamic role assignment using BTs, by defining the behavior of a leader agent that controls the roles in the organization and defining swap nodes that handle each role change.

BTs are very easy to implement, they are flexible and modular, which facilitates maintenance, scalability, and system updates. These characteristics prove to be a great solution for coordinating multiple robots, and being a great alternative to FSMs for defining \textbf{complex behaviors}.

In terms of performance, the use of the new BT-based coordination strategy enabled a statistically significant improvement in relation to the FSM-based system. When analyzing this improvement, it is important to consider that the developed BT is not a literal translation of the previous FSM, but a reinterpretation, taking advantage of the characteristics of the BTs. With this in mind, it is possible to state that the team's performance may have improved thanks to better control of the strategy's reactivity, as well as, for example, the better ability to define priorities between role changes in the organization.

%% file: contents/acknowledgments.tex
\section*{Acknowledgments}

We thank the \textanon{ThunderVolt}{\anontext} project members, for their help with the development of the solution and 
the \textanon{ThundeRatz robotics team, the Amigos da Poli Patrimonial Fund, and the Escola Politécnica from the Universidade de São Paulo}{\anontext} for all their support. In addition, the sponsorship of \textanon{STMicroelectronics, Altium, SolidWorks, and Circuibras to the ThunderVolt project.}{\anontext}.